\documentclass[aip,jcp,preprint,graphicx,nofootinbib]{revtex4-1}
\pdfoutput=1
\usepackage{graphicx}

\usepackage{algpseudocode}
\usepackage{textcomp}
\usepackage{comment}
\usepackage{array}
\newcolumntype{P}[1]{>{\centering\arraybackslash}p{#1}}

\includecomment{doubleblind}

\draft 

\newcommand*{\citen}[1]{%
  \begingroup
    \romannumeral-`\x 
    \setcitestyle{numbers}%
    \cite{#1}%
  \endgroup   
}

\begin{document}


\title{Exact Parallelization of the Stochastic Simulation Algorithm for Scalable Simulation of Large Biochemical Networks}

\def\sinaiAffiliation{Icahn Institute for Data Science and Genomic Technology, and Department of Genetics and Genomic Sciences, Icahn School of Medicine at Mount Sinai}
\author{Arthur P. Goldberg}
\email[Author to whom correspondence should be addressed: ]{Arthur.Goldberg@mssm.edu}
\affiliation{\sinaiAffiliation}
\author{David R. Jefferson}
\email{drjefferson@gmail.com}
\affiliation{Lawrence Livermore National Laboratory}
\author{John A. P. Sekar}
\email{karr@mssm.edu}
\affiliation{\sinaiAffiliation}
\author{Jonathan R. Karr}
\email{john.sekar@mssm.edu}
\affiliation{\sinaiAffiliation}

\date{\today}

\begin{abstract}
Comprehensive simulations of the entire biochemistry of cells have great potential to help physicians treat disease and help engineers design biological machines. But such simulations must model networks of millions of molecular species and reactions. 

The Stochastic Simulation Algorithm (SSA) is widely used for simulating biochemistry, especially systems with species populations small enough that discreteness and stochasticity play important roles.
However, existing serial SSA methods are prohibitively slow for comprehensive networks, and existing parallel SSA methods, which use periodic synchronization, sacrifice accuracy.

To enable fast, accurate, and scalable simulations of biochemistry, we present an exact parallel algorithm for SSA that partitions a biochemical network into many SSA processes that simulate in parallel.
Our parallel SSA algorithm exactly coordinates the interactions among these SSA processes and the species state they share by structuring the algorithm as a parallel discrete event simulation (DES) application and using an optimistic parallel DES simulator to synchronize the interactions.
We anticipate that our method will enable unprecedented biochemical simulations.
\end{abstract}


\maketitle 

\section{Introduction}
Models of biochemical systems play a critical role advancing medicine and bioengineering.
Technological advances in single-cell and genomic measurement are rapidly generating data that enable larger and more complex models of biochemical pathways and whole cells. 
But advances in the performance and accuracy of simulation techniques are needed for integrating models.

The Stochastic Simulation Algorithm (SSA) is a widely used method for predicting the time evolution of chemical systems transformed by chemical reactions.~\cite{gillespie1977exact,gillespie2007stochastic}
In particular, SSA can model the variability exhibited by chemical systems with small species populations,
conditions which often arise in small biological systems such as populations of individual cells.~\cite{maier2011quantification,yu2006probing}

\begin{doubleblind}
Our interest in creating tools to enable dynamical models of large biochemical systems has motivated us to parallelize SSA.
We are especially interested in whole-cell modeling, which creates genome-scale models of the known biochemical pathways in individual cells.~\cite{goldberg2018emerging,szigeti2018blueprint,karr2012whole,goldberg2016toward}
A whole-cell model of a cell with a large genome, like a human cell, contains tens of thousands of species and tens of thousands of reactions.
In addition, it contains numerous compartments that represent intracellular organelles such as the nucleus, mitochondria, lysosomes and others.
Whole-cell models are typically integrated for an entire cell cycle.~\cite{karr2012whole}
The complexity of whole-cell models and the duration of their integrations cause their simulations run slowly.
A highly parallel SSA algorithm could speed up simulations of models of large biochemical systems like whole-cell models.
\end{doubleblind}

There are multiple opportunities for parallelism in the simulation of biochemical networks due to the physical structure of cells, as well as due to artifacts of the limited ability of researchers to estimate their structure.
First, the compartmentalization of cells into organelles means that the biochemical networks of a cell decomposes into a set of biochemical sub-networks of the organelles, weakly connected via edges that represent a comparatively small number of exchange reactions.
Second, the subcompartmentalization of organelles into spatial domains such as chromosomal regions and individual mRNA means that the sub-networks of the organelles decompose into weakly connected cliques.
Third, the chemical specificity of most enzymes means that the sub-networks of the organelles further decompose into cliques. Fourth, our limited knowledge of the interactions among cellular processes means that our in silico biochemical networks have higher clustering coefficients than real biochemical networks. 

Exact implementations of SSA are computationally expensive when modeling systems that involve many reaction executions because they execute only one reaction per iteration.
In addition, since SSA models stochastic behavior, multiple runs are needed to obtain distributions and moments of predicted properties.

While extensive effort has been devoted to improving the performance of SSA \cite{thanh2014efficient,gillespie2003improved,auger2006r,gillespie2001approximate} an exact, parallel SSA algorithm that can concurrently leverage numerous processors to speed up an SSA simulation is not available.

We introduce a novel parallel algorithm for speeding up SSA.
Our approach conceives of SSA as a discrete event simulation (DES) and leverages optimistic approaches for parallel DES (PDES).~\cite{Fujimoto2000ParallelAD,Carothers2000,carothers2010deciding,jefferson1987time,jefferson2017virtual,mikida2016towards}
In contrast to prior work, our parallel SSA algorithm takes advantage of more of the parallelism inherent in large reaction networks, exactly synchronizes changes to the species population state that is shared among components of the algorithm, and, assuming that random number generators are properly managed, can exactly reproduce the predictions of a sequential SSA algorithm.

\section{The Stochastic Simulation Algorithm}
\subsection{A summary of SSA}
\label{A summary of SSA}
SSA solves the following problem. Consider a well-mixed container of chemical species, the chemical reactions that transform them, and a rate law for each reaction that provides its execution rate as a function of the species populations.
Given an initial state of this system, one can model the dynamic probability that it occupies any population state by representing state transition probabilities as a system of coupled differential equations.
Although this formulation, known as the Chemical Master Equation, is exact, it is not tractable for biologically interesting models because the state space is prohibitively large.

The Stochastic Simulation Algorithm models the dynamical behavior of this system by computing a sequence of reaction executions, choosing reactions and execution times so that the probability of generating a given trajectory equals the probability that would be provided from a solution of the Chemical Master Equation.

An algorithm for SSA, known as the Direct Method, was provided by Gillespie.~\cite{gillespie1977exact}
{\small
\begin{algorithmic}[1]
\Procedure{\textit{SSA Direct Method}}{}
  \State Set initial species populations
  \State $t \gets 0$ \Comment{simulation time}
  \State $\tau, \mu  \gets $ \textsc{Plan next reaction execution}
  \While{$t + \tau \leq$ max simulation time}
    \State $t \gets t + \tau$
    \State Update species populations according to reaction $\mu$
    \State $\tau, \mu  \gets $ \textsc{Plan next reaction execution}
  \EndWhile
\EndProcedure

\vspace{1mm}
\Function{Plan next reaction execution}{}
\For{r in reactions}
  \State Compute the rate (propensity) for $r$, $a_r$
\EndFor
\State $a_0 \gets \sum_r a_r$ \Comment{the total propensity}
\State \Comment{let $U()$ randomly sample the standard uniform distribution}
\State $\tau \gets (1/a_0) \ln(1/ U())$ \Comment{$\tau$ is the time the next reaction executes}
\State Choose $\mu$ by sampling $P[\mu] = a_\mu / a_0$ \Comment{$\mu$ is the next reaction}
\State return $\tau, \mu$
\EndFunction
\end{algorithmic}}

Detailed derivations for the Direct Method  \cite{gillespie1977exact,gibson2000efficient,cao2005multiscale,gillespie2007stochastic,anderson2007modified} rely on a key assumption of the Direct Method---that propensities $a_r$ remain constant in between reaction executions (equation (1) in \citen{anderson2007modified}).

We use the Direct Method for pedagogical purposes in this paper, but base our parallel SSA algorithm on an optimized SSA algorithm, the Next Reaction Method.~\cite{gibson2000efficient}


\algrenewcommand\algorithmicprocedure{\textbf{Event method}}
\section{Introduction to parallel SSA}
This section provides an overview of the parallel SSA algorithm.
\subsection{Objectives}
The parallel SSA algorithm we present has two primary objectives.
\begin{enumerate}
  \item Speed up the performance of SSA on a large reaction network by partitioning the network into multiple sub-networks and integrating the sub-networks in parallel.
  \item Exactly simulate reaction networks, so that the results of a parallel simulation are the same as the results of a sequential simulation. This objective is satisfied by the parallel SSA algorithm presented below.
\end{enumerate}
\subsection{Opportunities for parallelism}
A reaction network in a large biochemical model like a whole-cell model provides two types of opportunities for parallelism.
First, the sub-networks of reactions in separate compartments may be sufficiently decoupled from each other that they can be simulated in parallel.
The amount of a compartment's coupling is measured by the mean sum of the rates of its reactions that exchange species with other compartments, relative to the mean sum of the rates of all reactions in the compartment.
Multiple prior studies have taken this approach to parallelize SSA.~\cite{jeschke2008parallel,mazza2012relevance,dematte2008parallel,wang2009experimental,hallock2014simulation}
Second, reactions within a compartment may also be partitioned into sub-networks that are sufficiently decoupled that they can be simulated in parallel. 

These two types of potential parallelism are complementary and can be combined in a single parallelization.
Our parallel SSA approach below abstracts away the distinctions between them and simply partitions a reaction network into sub-networks.
Nevertheless, we note these opportunities for parallelism because prior work has leveraged the first one, and this discussion may help readers develop intuition for parallel SSA.

\subsection{Overview of the parallel SSA approach}
To provide an overview of the parallel SSA algorithm we briefly summarize its high-level steps, broken down into its parallelization and simulation phases.
\vspace{5mm}\\
\textbf{Parallelize}
\begin{enumerate}
  \item Read a model to simulate, including its reactions and rate laws.
  \item Generate a graph $G$ that characterizes the dependencies among reactions and species in the model.
  \item Partition $G$ into sub-networks of reactions that will be simulated in parallel.
  \item Identify the species that are shared between multiple sub-networks of reactions, and partition them.
\end{enumerate}
\textbf{Simulate}
\begin{enumerate}
  \item Read a simulation configuration for the model, including the initial populations of species and other initial conditions, and a maximum simulation time.
  \item Instantiate an optimistic object-oriented (OO) parallel discrete event simulation (PDES) environment on a parallel computer.~\cite{Jefferson1985,Carothers2000,Barnes2013,carothers2010deciding,carothers1999efficient,jefferson1987time,jefferson2017virtual,schordan2015reverse,mikida2016towards,mikida2018adaptive}
  \item Map each reaction sub-network of $G$ onto an SSA simulation object in the optimistic PDES environment, instantiate the object, and send it an initialization event message.
  \item Map each set of shared species onto a Shared Species Population (SSP) simulation object in the PDES environment, instantiate the object and send it an initialization event message.
  \item Initialize all other objects in the simulation.
  \item Run the parallel SSA simulation.
\end{enumerate}
These steps and components are thoroughly described below.

\section{The parallel SSA algorithm}
This section presents the concepts underlying the parallel SSA algorithm.
\subsection{Transforming a reaction network into a parallel simulation}

\subsubsection{Encode the reaction network's dependencies into a directed graph}
\label{Encode the reaction network's dependencies in a directed graph}
\begin{figure}[!htbp]
  \centering
  \includegraphics[width=.7\linewidth]{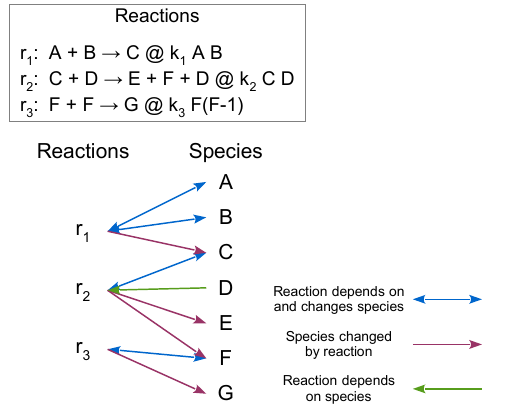}
\caption{Relationships between reactions and species encoded in dependency graph $G$.
\textmd{The relationships between reactions and species are encoded in a directed graph whose nodes are reactions and species.
Directed edges indicate data dependencies.
An edge from a reaction and its rate law to a species indicates that executing the reaction changes the species' population.
An edge from a species to a reaction indicates that the reaction's rate law uses the species' population.
For example, $G$ contains an edge from enzyme D to reaction $r_2$ because the rate law for $r_2$ is a function of the population of enzyme D.
Bi-directional edges indicate both of these dependencies.\\
The @ signs in the list of reactions separate each reaction from its rate law, a mathematical expression that computes the rate at which the reaction executes.
In a rate law a species symbol represents the number of molecules of the species in the container being simulated.}}
\label{fig:bipartite_dependencies}
\end{figure}

Let $R$ and $S$ represent all reactions and species, and $r_i$, $s_j$, and $l_k$ represent individual reactions, species and rate laws, respectively.
We map the dependencies between these components onto a bipartite directed graph $G(v, e)$, where $v = R \bigcup S$.
$G$ is constructed as follows.
A directed edge $(s_j, r_i)$ is added to $G$ if species $s_j$ participates in rate law $l_i$, thereby encoding the  dependency of reaction $r_i$ on species $s_j$.
Similarly, a directed edge $(r_i, s_j)$ is added to $G$ if $s_j$ has a non-zero stoichiometry
in reaction $r_i$, thereby encoding the dependency of $s_j$ on reaction $r_i$ that arises because executing reaction $r_i$ changes the populations of species $s_j$.
Figure~\ref{fig:bipartite_dependencies} illustrates $G$ for a small example network.

\subsubsection{Partition the dependency graph}
\label{Partition the dependency graph}
To identify reaction network components that can be simulated in parallel, we partition $G$ into two types of sub-networks.
The first type of sub-network contains reactions, and all of the species that participate only in reactions in the sub-network\footnote{More precisely, these are the species connected in $G$ to reactions in the sub-network and not connected to any other reactions.}
(e.g., see partitions $\alpha$ and $\beta$ in in Figure~\ref{fig:dependency_partition}).
The species contained in a reaction sub-network are called \textit{local} species, because they're used only by the reactions in their local sub-network.
For example, in Figure~\ref{fig:dependency_partition} species $A$ and $B$ are local to sub-network $\alpha$.

The second type of sub-network created by a partition contains only \textit{shared} species, which are used by reactions in two or more reaction sub-networks.

\subsubsection{Designing the partitioning algorithm}
The goal of the partitioning algorithm is to identify partitions into sub-networks that can be rapidly simulated in parallel.
In the near-term, we plan to use linkage clustering algorithms to partition the dependency graph $G$.
Intuitively, this will minimize the number of state changes that would need to be synchronized between the DES objects which are encoded into the sub-networks.

Long-term, we aim to develop an algorithm for identifying the optimal partitioning that minimizes the expected run-time on a parallel computer.
Due to the challenges to analyzing the expected run-time of a parallel SSA simulation, the optimal partitioning will likely need be to identify empirically by testing the performance of putative partitionings.





\subsubsection{Encoding the reaction network into simulation objects}
The two types of sub-networks produced by a partition are mapped into two types of DES objects.

\begin{figure}[!htbp]
  \centering
  \includegraphics[width=0.7\linewidth]{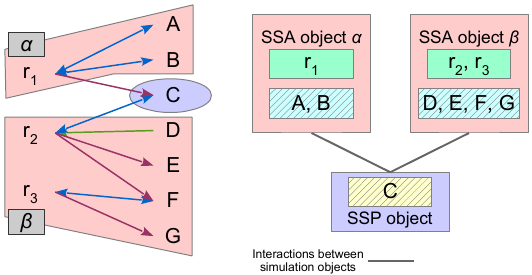}
  \caption{A partition of the dependencies illustrated in Figure~\ref{fig:bipartite_dependencies}.
\textmd{The dependency graph $G$ in Figure~\ref{fig:bipartite_dependencies} is partitioned into two reaction sub-networks, $\alpha$ and $\beta$, and 1 shared species sub-network.
These are then mapped onto the SSA objects $\alpha$ and $\beta$, and an SSP object.
In this small network each SSA object interacts with the SSP object, but in reaction networks suitable for parallel execution by parallel SSA, each SSA object would interact with a small fraction of the  SSP objects, and \textit{vice versa}.}}
\label{fig:dependency_partition}
\end{figure}

Each sub-network of reactions defines the reactions that are integrated by one OO DES object that simulates SSA, called an \textit{SSA object}.
For example, in Figure~\ref{fig:dependency_partition} SSA object $\alpha$ integrates reaction $r_1$ and stores local species $A$ and $B$, while SSA object $\beta$ integrates reactions $r_2$ and $r_3$ and stores local species $D$, $E$, $F$, and $G$.

Each sub-network of shared species is mapped into one OO DES Shared Species Population (SSP) object which will coordinated the populations of the shared species during a simulation.
For example, in Figure~\ref{fig:dependency_partition} shared species $C$ is stored in an SSP object and is used by reactions $r_1$ and $r_2$, which are executed by two different SSA objects, $\alpha$ and $\beta$, respectively.

\subsection{Architecture of parallel SSA}

This parallel SSA algorithm has been designed to run as a parallel OO DES application on top of an optimistic PDES simulator.~\cite{Jefferson1985,Barnes2013,Carothers2000,Fujimoto2000ParallelAD,fujimoto1990parallel,mikida2016towards,jefferson1987time,jefferson2017virtual,schordan2015reverse}
The major advantage of this approach is that the optimistic PDES simulator will be responsible for running the objects in the parallel SSA algorithm in simulation time order.
The major types of simulation objects in parallel SSA will be the SSA and SSP classes defined above.
In addition, other utility classes will be added, such as classes for initialization and for checkpointing model predictions.


\section{The parallel SSA algorithm}
\label{section:The parallel SSA algorithm}
This section presents our approach for an exact parallel SSA algorithm.
We begin by restructuring SSA as an OO DES application and defining the structure of DES application classes.
In sub-section~\ref{section:An exact parallel SSA algorithm} we illustrate the features and components of the algorithm, which we follow with a sub-section containing pseudocode for its classes.
The final sub-section describes how the algorithm handles compartments. 

\subsection{Structure SSA as an object-oriented discrete event simulation application}
\label{Structure SSA as an object-oriented discrete event simulation application}
Many variations of the SSA algorithm have been published, beginning with two in Gillespie's original paper.~\cite{gillespie1977exact}
SSA variations have the characteristics of a DES application.
First, their events, which simulate reaction executions, occur at a discrete time instants.
And, second, these events are dynamically scheduled by computations at earlier simulation times.
But published SSA variations do not use DES because they're written as free-standing algorithms.

Since an optimistic parallel OO DES simulator will be used to synchronize the parallel SSA algorithm, we begin the discussion of synchronization by recasting the Direct Method (presented in section~\ref{A summary of SSA}) as an OO DES application.
This implementation of the Direct Method OO DES class executes two event methods, \textsc{\textit{Initialize Direct Method}} and \textsc{\textit{Execute and schedule reaction}}:

{\small
\begin{algorithmic}[1]
\Procedure{\textit{Initialize Direct Method}}{\textit{0, self, self, initial populations, max simulation time}}
  \State Set initial species populations
  \State $t_m \gets $ \textit{max simulation time}
  \State \Comment{Call Direct Method function defined in section~\ref{A summary of SSA}}
  \State $\tau, \mu  \gets $ \textsc{Plan next reaction execution()}
  \State send \textit{Execute and schedule reaction($\tau$, self, self, $\mu$)}
\EndProcedure

\vspace{1mm}
\Procedure{\textit{Execute and schedule reaction}}{\textit{$t$, self, self, $\mu$}}
  \If{$t \leq t_m$}
    \State Update species populations according to reaction $\mu$
    \State $\tau, \mu  \gets $ \textsc{Plan next reaction execution()}
    \State send \textit{Execute and schedule reaction($t + \tau$, self, self, $\mu$)}
  \EndIf
\EndProcedure
\end{algorithmic}}

All OO DES application objects interact exclusively through event messages, a strict interface that makes it possible for parallel OO DES simulators to execute them.

As illustrated by this pseudo-code for the Direct Method class, all DES event messages in this paper have the form

\textit{Message type(time, sender, receiver, [arguments])},\\
where \textit{Message type} is the type of the message, \textit{time} is the simulation time at which the message will be received and executed, \textit{sender} identifies the simulation object that sends the message, \textit{receiver} identifies the simulation object that will receive the message, and the optional \textit{arguments} are data carried by the message from its \textit{sender} to its \textit{receiver}.

The algorithm for a DES application class is defined by its message handler event methods, whose names and signatures must exactly match the types and fields of messages received by the class, although the handler method names are written in \textsc{small caps} to distinguish them from the messages.
Throughout the execution of a handler the simulation time of the \textit{receiver} object handling an event message is automatically  \textit{time}.


\subsection{An exact parallel SSA algorithm}
\label{section:An exact parallel SSA algorithm}
This section presents an algorithm that focuses on the key logic needed to exactly parallelize SSA.
Performance optimizations for the algorithm follow in section \ref{section:Optimizing parallel SSA}.

The exact parallel SSA algorithm is a parallel OO DES application that executes two class types, SSA objects and Shared Species Population (SSP) objects.
Updates and reads of local species stored in SSA objects always occur at the correct simulation times, so we focus on shared species.
To exactly synchronize shared species, updates and reads of their populations must also occur at the correct times.
More specifically, all updates to shared species populations by a reaction must change the populations in SSP objects at the time the reaction executes.
And when an SSA object uses a shared species at time $t$ it must obtain the SSP's value for the species' population at $t$.

Correct timing of updates and reads of shared species populations is achieved by explicitly accessing shared species at SSP objects.
When a reaction executed by an SSA object updates the population of a shared species, the object updates the population at the SSP by sending an \textit{Adjust populations} message to the SSP.
And the SSP handles an \textit{Adjust populations} message by sending a zero-delay \textit{Populations} message to each SSA object that uses the shared species which were updated in the \textit{Adjust populations} message.

\begin{figure*}[!htbp]
  \centering
  \includegraphics[width=\linewidth]{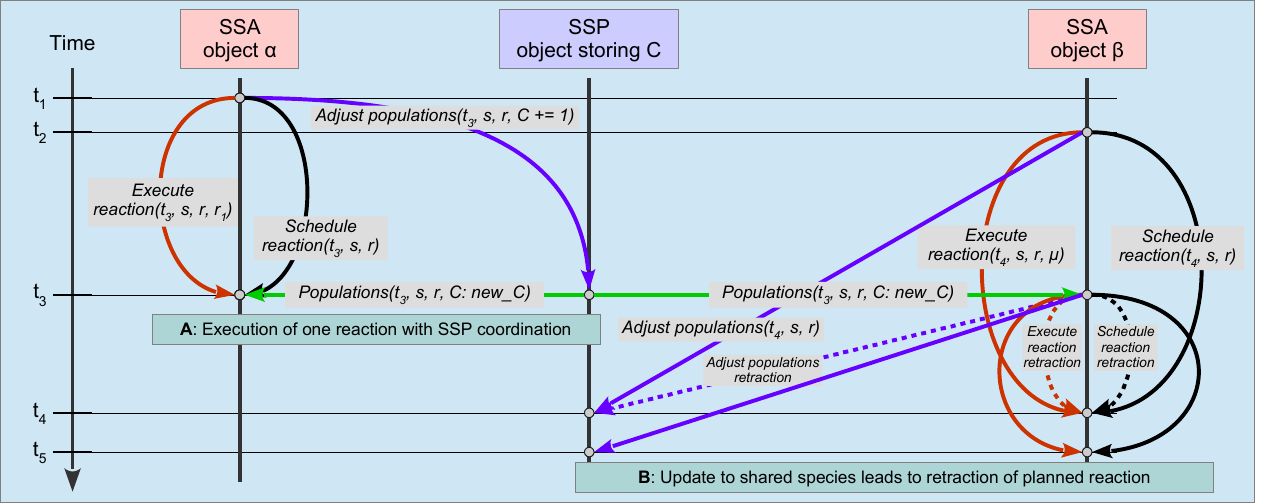}
\caption{Example event message interactions between SSA and SSP objects.
\textmd{To conserve space, the event message fields \textit{time, sender, receiver} are abbreviated \textit{$t_i$, s, r}.
\textbf{A}: The interactions between SSA object $\alpha$ and the SSP show the messages involved in a single reaction execution coordinated with the SSP, as described in section~\ref{section:An exact parallel SSA algorithm}.
\textbf{B}: The \textit{Populations} message that updates the population of shared species C at SSA object $\beta$ at time $t_3$ causes SSA $\beta$ to cancel its planned execution of reaction $\mu$ at time $t_4$, as discussed in section \ref{section:Handling updates to shared species used by rate laws}. Each \textit{retraction} message is indicated by a dashed arrow that has the same color as the message being retracted, and points to the same object at the same time as it does.}}
\label{fig:messages_normal_and_cancellation}
\end{figure*}

Table~\ref{tab:message_types} lists these messages and Figure~\ref{fig:messages_normal_and_cancellation} illustrates their dynamics.
The interactions between SSA object $\alpha$ and the SSP over the time interval $t_1$ to $t_3$ (see section \textbf{A} of Figure~\ref{fig:messages_normal_and_cancellation}) execute one reaction, $r_1$.
At time $t_1$ three event messages are sent to schedule the reaction and the parallel coordination it requires:

\begin{table*}[!htbp]
\caption{Event message types used by Parallel SSA}
\label{tab:message_types}

\begin{tabular}{  p{2.3cm}  P{1.7cm}  P{2cm}  p{4cm}  p{6cm} }
 \hline
\textbf{Message type} & \textbf{Sender object type} & \textbf{Receiver object type} & \textbf{Arguments} (not including the required arguments \textit{time}, \textit{sender}, and \textit{receiver}) & \textbf{Parallel SSA semantics and receiver object action}\\
\hline 
\textit{Execute reaction} & SSA & SSA & $\mu$ & Execute reaction $\mu$\\
\hline 
\textit{Schedule reaction} & SSA & SSA & & Schedule the next reaction \\ 
\hline 
\textit{Adjust populations} & SSA & SSP & \textit{pop\_changes} & Convey the stoichiometry of a reaction's shared species to the SSP, which updates their populations\\
\hline 
\textit{Populations} & SSP & SSA & \textit{shared\_species\_pops} & Provide populations of the requested species to the SSA\\ 
\hline 
\end{tabular}
\end{table*}

\begin{itemize}
    \item \textit{Execute reaction} schedules the execution of $r_1$ at SSA $\alpha$ at $t_3$. Every reaction execution requires this message.
    \item \textit{Schedule reaction} causes SSA $\alpha$ to schedule the next reaction. Every reaction execution requires this message.
    \item \textit{Adjust populations} schedules the SSP to increment the population of C  at $t_3$, as per the stoichiometry of reaction $r_1$.
    Every reaction execution that changes the population of shared species requires this message.
\end{itemize}
At time $t_3$, in response to the \textit{Adjust populations} message, the SSP sends a \textit{Populations} message that contains the shared populations to SSA $\alpha$.

The algorithms that send and handle these messages are detailed in sections \ref{section:The SSP class} and \ref{section:The SSA class}.

\subsubsection{Handling updates to shared species used by rate laws}
\label{section:Handling updates to shared species used by rate laws}
A fundamental assumption SSA is that the propensities\footnote{The standard notation $a_k (X(t))$ indicates the propensity of reaction $k$---computed using $k$'s rate law---as a function of the species populations at time $t$, $X(t)$.} $a_k (X(t))$ are constant except when reaction executions change species counts (equation (1) in.~\cite{anderson2007modified})\footnote{Some SSA algorithms model time-varying propensities, for example to simulate containers with varying volume, as in section V of \cite{anderson2007modified}.}
Parallel SSA revises this assumption by adding that \textit{\textbf{propensities can also change when the populations of shared species used by propensities are updated}}.
Changes external to an SSA object can thus invalidate the computation that scheduled a reaction execution.

Figure~\ref{fig:messages_normal_and_cancellation} illustrates this situation in the interactions among 2 SSA objects and an SSP object.
Focus on the SSP and SSA object $\beta$ over the time interval $t_2$ to $t_5$ (section \textbf{B} of Figure~\ref{fig:messages_normal_and_cancellation}):
\begin{enumerate}
    \item At time $t_3$, in response to the \textit{Adjust populations} it received from SSA $\alpha$, the SSP sends a \textit{Populations} message that updates the population of shared species C at SSA $\beta$.
    \item At $t_3$ SSA $\beta$ receives the message, updates the population of species C, determines that the population of C was used by the propensity calculations made when scheduling its pending reaction $\mu$, and concludes that the assumption made when scheduling $\mu$---that the propensities calculated at time $t_2$ would remain constant until time $t_4$---no longer holds. Thus, SSA $\beta$ must now cancel the scheduled execution of reaction $\mu$, and reschedule its next reaction.
    \item SSA $\beta$ cancels each of the messages it sent at $t_2$ to schedule the execution of reaction $\mu$ by sending a corresponding \textit{retraction} message at $t_3$ (see the three messages labeled \textit{retraction} and indicated by dashed arrows in Figure~\ref{fig:messages_normal_and_cancellation}).
    \textit{Retraction} messages \cite{lomow1991mechanisms} are application messages that completely cancel the effects of a corresponding retracted message, including the possible execution of the retracted message and that execution's side effects.
    They are a fully developed feature of the Virtual Time .~\cite{Jefferson1985,jefferson2017virtual,jefferson2020virtual_time_III} optimistic parallel synchronization theory which is implemented by Time Warp optimistic PDES simulators \cite{jefferson1987time,Carothers2000}
    \item SSA $\beta$ now reschedules its next reaction, which Figure~\ref{fig:messages_normal_and_cancellation} illustrates as executing at time $t_5$.
\end{enumerate}

The algorithms for the parallel SSA classes that implement these actions are detailed in the next section.

\subsection{Class algorithms for parallel SSA}
\label{section:Class algorithms for parallel SSA}
This section presents event methods for the SSA and SSP classes that implement the parallel SSA algorithm.
To simplify this section's presentation all shared species are stored in a single SSP object, and we consider a system with only one biological compartment, corresponding to a well-mixed container in the stochastic simulation's analytic framework .~\cite{gillespie1977exact}
These limitations are relaxed  in section~\ref{section:Optimizing parallel SSA} below.

\subsubsection{The SSP class}
\label{section:The SSP class}
An SSP object executes two types of events: \textsc{\textit{Initialize}} initializes the SSP, and \textsc{\textit{Adjust populations}} updates the populations of specified shared species.

{\small
\begin{algorithmic}[1]
\Procedure{\textit{Initialize}}{0, self, self, initial\_populations, species\_locs}
\State \Comment{initialize initial populations of all shared species}
\State self.populations = initial\_populations
\State self.SSA\_map = species\_locs \Comment{map shared species to their SSAs}
\EndProcedure
\end{algorithmic}}

{\small
\begin{algorithmic}[1]
\Procedure{\textit{Adjust populations}}{time, SSA\_obj, self, pop\_changes}
\State add the changes in pop\_changes to self.populations
\State \Comment{send updated populations to SSAs that depend on them}
\For{SSA in self.SSA\_map}
    \State shared\_species\_pops = empty dictionary
    \For{species in pop\_changes}
        \If{species in self.SSA\_map[SSA]}
            \State shared\_species\_pops[species] = self.populations[species]
        \EndIf
    \EndFor
    \If{shared\_species\_pops}
        \State send \textit{Populations}(time, self, SSA, shared\_species\_pops)
    \EndIf
\EndFor
\EndProcedure
\end{algorithmic}}
{\small \textsc{\textit{Adjust populations}}} adjusts the populations of shared species in {\small self.populations} by the changes in {\small pop\_changes}.
In lines 4--14 the SSP acts like a write-through cache, forwarding updated shared species population values to the SSAs executing reactions that use the species in their rate laws.


\subsubsection{The SSA class}
\label{section:The SSA class}
This section defines the algorithms used by the SSA class.
We adapt Gibson and Bruck's Next Reaction Method (NRM) \cite{gibson2000efficient,anderson2007modified} to determine the reactions that execute and their execution times.
However, to simplify the presentation we do not incorporate NRM's optimization that avoids recomputing propensities which do not depend on changed species populations.
Instead, in section~\ref{section:Optimizing parallel SSA} we identify it as an optimization that should be added to the algorithm.


An SSA object executes four types of event methods.
They correspond 1-to-1 to the messages illustrated in Figure~\ref{fig:messages_normal_and_cancellation} and described in section~\ref{section:An exact parallel SSA algorithm}: \textsc{\textit{Initialize}} initializes the object\footnote{Note that the \textsc{\textit{Execute and schedule reaction}} event method used by the SSA class presented in section~\ref{Structure SSA as an object-oriented discrete event simulation application} has been decomposed into two event methods, which is required by the reasoning at the end of section~\ref{Dependencies among simultaneous parallel SSA event messages}.}, \textsc{\textit{Schedule reaction}} schedules the next reaction, \textsc{\textit{Execute reaction}} executes a reaction, and \textsc{\textit{Populations}} receives updated values for the populations of the shared species used by the SSA object.

{\small
\begin{algorithmic}[1]
\Procedure{\textit{Initialize}}{0, self, self, reactions, initial\_populations}
  \State self.reactions = reactions \Comment{reactions and their rate laws}
  \State self.local\_species = initial\_populations[\textquotesingle local\_species\textquotesingle]
  \State self.cached\_shared\_species = initial\_populations[\textquotesingle shared\_species\textquotesingle]
  \State self.SSP = the single SSP
  \State self.next\_rxn.time = 0    \Comment{self.next\_rxn tracks a pending reaction}
  \State self.next\_rxn.reaction = $\emptyset$
  \State $\mu, \tau_\mu =$ \textit{\textsc{Select initial reaction}}(self)
  \State \textit{\textsc{Send reaction events}}($\mu$, $\tau _ \mu$, self)
\EndProcedure
\end{algorithmic}}
\textit{\textsc{Initialize}} only runs once, at time 0.

{\small
\begin{algorithmic}[1]
\Function{Select initial reaction}{self}
    \State \Comment{Select the first reaction, saving propensities and times}
    \For{$k$ in self.reactions}
        \State self.$a_k$ = $a_k$ = propensity for $k$
        \State $r_k$ = uniform(0, 1) random number
        \State self.$\tau_k = \tau_k = (1 / a_k) \ln(1/r_k)$ \Comment{execution time for $k$}
    \EndFor
    \State $\mu = k$ that satisfies $\min_k\{\tau_k\}$ \Comment{Select next reaction}
    \State \textbf{return} $\mu, \tau_\mu$
\EndFunction
\end{algorithmic}}

To support the event-driven model of OO DES, the reaction scheduling code must be separated into a different function for each event.
Each function determines and returns the next reaction to execute, $\mu$, and the time when it will execute, $\tau_\mu$.
The first such function, \textsc{Select initial reaction}, which is called by \textit{\textsc{Initialize}}, is mathematically equivalent to the initialization of NRM (lines 2--5 of Alg. 2 in \citen{anderson2007modified}).

{\small
\begin{algorithmic}[1]
\Function{Send reaction events}{$\mu$, $\tau_\mu$, self}
  \State send \textit{Execute reaction}($\tau_\mu$, self, self, $\mu$)
  \State send \textit{Schedule reaction}($\tau_\mu$, self, self)
  \State species\_changes = shared species with stoichiometry $\neq 0$ in $\mu$
  \If{species\_changes}
    \State send \textit{Adjust populations}($\tau_\mu$, self, self.SSP, species\_changes)
  \EndIf
  \State self.next\_rxn.reaction = $\mu$
  \State self.next\_rxn.time = $\tau_\mu$
\EndFunction
\end{algorithmic}}

The \textsc{Send reaction events} function sends all of the event messages transmitted by in SSA object.
It is called by each of the three event methods that send messages.
Each call sends all the messages associated with the execution of one reaction.
Lines 11--12 record the reaction and its execution time for later use.

{\small
\begin{algorithmic}[1]
\Procedure{\textit{Schedule reaction}}{t, SSA\_obj, self}
  \State \Comment{Schedule a reaction after executing one}
  \State $\mu$ = self.next\_rxn.reaction
  \State $\mu, \tau_\mu =$ \textit{\textsc{Select after reaction execution}}(self, t, $\mu$)
  \State \textit{\textsc{Send reaction events}}($\mu$, $\tau _ \mu$, self)
\EndProcedure
\vspace{2mm}
\Function{Select after reaction execution}{self, $t$, $\mu$}
    \State \Comment{Select next reaction after executing $\mu$; update self.$a_k$ and self.$\tau_k$}
    \State $s =$ self
    \For{$k$ in self.reactions}
        \State $\overline{a}_k$ = new propensity for $k$
        \If{$k \neq \mu$}
            \State self.$\tau_k = \tau_k = \frac{s.a_k}{\overline{a}_k}(s.\tau_k - t) + t$
        \EndIf
    \EndFor
    \State $r_\mu$ = uniform(0, 1) random number
    \State self.$\tau_\mu = \tau_\mu = (1 /~\overline{a}_\mu) \ln(1/r_\mu)$
    \State $\mu = k$ that satisfies $\min_k\{\tau_k\}$ \Comment{Select next reaction}
    \For{$k$ in self.reactions}
        \State self.$a_k = \overline{a}_k$
    \EndFor
    \State \textbf{return} $\mu, \tau_\mu$
\EndFunction
\end{algorithmic}}
\textsc{\textit{Schedule reaction}} implements a DES version of the body of the NRM's iterative loop.

\textsc{Select after reaction execution}, which selects the next reaction after a reaction executes, is mathematically equivalent to the code in NRM's iterative loop (lines 5 and 7--10 of Alg. 2 in \citen{anderson2007modified}).

{\small
\begin{algorithmic}[1]
\Procedure{\textit{Populations}}{t, SSP\_obj, self, shared\_species\_pops}
  \State self.cached\_shared\_species = shared\_species\_pops
  \If{time < self.next\_rxn.time}
    \State \Comment{Cancel next reaction and schedule again}
    \State \textit{\textsc{Cancel scheduled reaction}}(self)
    \State $\mu, \tau_\mu =$ \textit{\textsc{Select after reaction cancellation}}(self, t)
    \State \textit{\textsc{Send reaction events}}($\mu$, $\tau _ \mu$, self)
  \EndIf
\EndProcedure

\vspace{2mm}
\Function{Cancel scheduled reaction events}{self}
  \State \Comment{Cancel events related to the execution of reaction self.next\_rxn}
  \State $\mu$ = self.next\_rxn.reaction
  \State $\tau_\mu$ = self.next\_rxn.time
  \State retract \textit{Execute reaction}($\tau_\mu$, self, self, $\mu$)
  \State retract \textit{Schedule reaction}($\tau_\mu$, self, self)
  \State species\_changes = shared species with stoichiometry $\neq 0$ in $\mu$
  \If{species\_changes}
    \State retract \textit{Adjust populations}($\mu$, self, self.SSP, species\_changes)
  \EndIf
\EndFunction

\vspace{2mm}
\Function{Select after reaction cancellation}{self, $t$}
    \State \Comment{Select next reaction after cancellation, updating self's $a_k$ and $\tau_k$}
    \State $s =$ self
    \For{$k$ in self.reactions}
        \State $\overline{a}_k$ = new propensity for $k$
        \State self.$\tau_k = \tau_k = \frac{s.a_k}{\overline{a}_k}(s.\tau_k - t) + t$
    \EndFor
    \State $\mu = k$ that satisfies $\min_k\{\tau_k\}$ \Comment{Select next reaction}
    \For{$k$ in self.reactions}
        \State self.$a_k = \overline{a}_k$
    \EndFor
    \State \textbf{return} $\mu, \tau_\mu$
\EndFunction
\end{algorithmic}}

\textsc{\textit{Populations}} records updated populations for the shared species used by an SSA object's rate laws.
If a future reaction is pending then it must be cancelled and reaction scheduling must be redone, as previewed in section~\ref{section:Handling updates to shared species used by rate laws} above.
This novel aspect of the parallel SSA algorithm is handled by a pair of functions:
\textsc{Cancel scheduled reaction events} cancels all of the events related to the previously scheduled reaction, and \textsc{Select after reaction cancellation} then selects the next reaction to execute.
In the former function, each ``retract \textit{Event message}'' operation sends a \textit{retraction} message that cancels the previously sent \textit{Event message}.

\textsc{Select after reaction cancellation} schedules the next reaction.
We adapt the NRM to determine the reaction times.
Because \textit{no} reactions have executed since propensities were last calculated, all reactions---including the one that was cancelled---can be treated like the reactions that did not execute in the iterative NRM loop (see lines 4--9 of Alg. 2 in \citen{anderson2007modified}).
Thus, lines 27--31 of \textsc{Select after reaction cancellation} calculate a new propensity and execution time for each reaction, and select the reaction with the minimum time.
No random numbers are needed.

{\small
\begin{algorithmic}[1]
\Procedure{\textit{Execute reaction}}{t, SSA\_obj, self, $\mu$}
  \State update self.local\_species according to the stoichiometry of $\mu$
\EndProcedure
\end{algorithmic}}
\textsc{\textit{Execute reaction}} updates the local species populations according to the stoichiometry of the reaction being executed.

\subsection{Correctness of the parallel SSA algorithm}
The algorithms for the SSP and SSA class methods define an exact parallel SSA algorithm because they maintain these invariants:
\begin{description}
\item[Read timing] Propensity calculations read species populations at the correct times.
\item[Write timing] Updates to species populations are performed at the correct times.
\item[Static propensities] The propensities used to schedule every reaction that executes remain constant between the time they are computed and the reaction's execution.
\end{description}

We consider only shared species, since locally stored species are trivially read and updated at the correct times.
\textbf{Read timing} holds because the most recent populations of all shared species are used to calculate each propensity: whenever the SSP storing a shared species receives an \textit{Adjust populations} message it responds by sending a zero-delay \textit{Populations} message to each SSA object that uses the shared species which were updated in the \textit{Adjust populations} message.
\textbf{Write timing} holds because the populations of all shared species modified by a reaction are updated at the SSP via an \textit{Adjust populations} message that is executed at the time the reaction executes.
And \textbf{Static propensities} holds because whenever a species population used by a propensity calculation changes, the reaction that depends on the change is cancelled and scheduling is redone, as performed in lines 5--7 of \textsc{\textit{Populations}}.

However, one final issue must be resolved to complete this informal correctness proof.
The dependencies among \textit{simultaneous} SSA event messages need to be addressed.
The next sub-section defines these dependencies and describes how the parallel SSA algorithm ensures that simultaneous event messages execute in the correct order.

\subsubsection{Dependencies among simultaneous parallel SSA event messages}
\label{Dependencies among simultaneous parallel SSA event messages}
When a reaction executes all four event messages associated with the reaction execute at the simulation time of the reaction (see section \textbf{A} of Figure~\ref{fig:messages_normal_and_cancellation}).
To achieve correctness parallel SSA must control the execution order of simultaneous event messages.
Specifically, they should be executed in an order consistent with the logic of sequential SSA.
This section presents the rationale for that order, and the simulation mechanisms that achieve the order.

\begin{figure}[!htbp]
  \centering
  \includegraphics[width=0.7\linewidth]{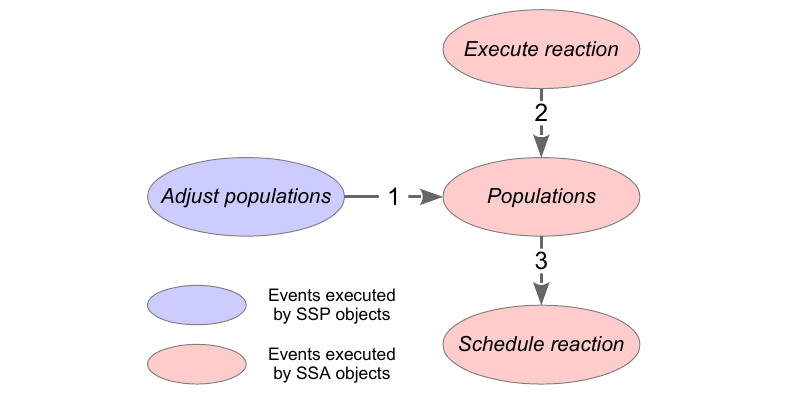}
\caption{Dependencies among \textit{simultaneous} parallel SSA events associated with a single reaction execution.
\textmd{A directed edge from event message \textit{x} to event message \textit{y} means that 
the event that executes message \textit{x} must occur before the event that executes message \textit{y}.
The rationale for each directed edge is provided in the text.}}
\label{fig:event-timing-dependencies}
\end{figure}

Figure~\ref{fig:event-timing-dependencies} illustrates logical dependency relationships between \textit{simultaneous} parallel SSA event messages.
These reasons explain the dependencies in Figure~\ref{fig:event-timing-dependencies}:
\begin{enumerate}
\item \textit{Populations} is a response to \textit{Adjust populations}, so \textit{Adjust populations} must execute before \textit{Populations}.
\item \textit{Populations} must follow \textit{Execute reaction} so that species populations are not altered before \textit{Execute reaction} is executed.
\item \textit{Populations} must precede \textit{Schedule reaction} so that \textit{Schedule reaction} sees fully updated populations, as occurs in sequential SSA algorithms.
\end{enumerate}

Parallel SSA achieves the ordering presented in Figure~\ref{fig:event-timing-dependencies} in two levels.
First, globally at any instant of time SSP objects must execute before SSA objects, as the alternative would be inconsistent with the dependencies in Figure~\ref{fig:event-timing-dependencies}.\footnote{All simultaneous messages received by an object at a given simulation time must be passed to the object as an \textit{event message set}.~\cite{jefferson2020virtual_time_III}
Therefore, SSPs must precede SSAs at a given time to ensure that \textit{Adjust populations} executes before \textit{Populations}.}
And second, locally within each object type, messages must execute in the fully determined order shown in Figure~\ref{fig:event-timing-dependencies}.

Two standard PDES mechanisms can implement these two levels of ordering.
Globally, simultaneous events at different PDES objects can be ordered by controlling a sub-time of the simulation time.
SSP objects can be forced to execute before SSA objects when they execute simultaneously by always giving SSP objects a smaller sub-time than SSA objects.
Locally, within a single simulation object the execution of simultaneous event messages must be ordered.
To implement the ordering above SSA objects must execute their events in this order: \textit{Execute reaction}, \textit{Populations}, \textit{Schedule reaction}.
This ordering is straightforward to achieve.
\begin{doubleblind}
For example, a sequential OO DES engine called \textit{DE Sim} that we wrote provides declarative support for controlling the execution order of simultaneous events.~\cite{goldberg2020DEsim}
\end{doubleblind}

The requirement that \textit{Populations} executes between \textit{Execute reaction} and \textit{Schedule reaction} forces parallel SSA to logically distinguish the event that executes a reaction from the event that schedules the next reaction.

\subsection{Compartments}
\label{section:Compartments}
Multiple compartments are fully supported by the parallel SSA algorithm.
Consider a system with two adjacent compartments, $c_1$ and $c_2$, and a single exchange reaction $x$ that transfers a species from $c_1$ to $c_2$.
Letting the species being transferred be named $s[c_1]$ and $s[c_2]$ in $c_1$ and $c_2$, respectively, reaction $x$ can simply be $s[c_1] \rightarrow s[c_2]$.
Let the reactions contained in $c_1$ and $c_2$ be mapped to SSA objects $S_1$ and $S_2$ respectively, with $S_1$ executing reaction $x$ and $s[c_2]$ a shared species used by both $S_1$ and $S_2$.  
When $x$ executes, the population of $s$ is decremented in $c_1$ and incremented in $c_2$.

This exchange reaction is naturally handled by the parallel SSA algorithm.
Since $S_1$ executes $x$, the population change in $S_1$ occurs at the time $x$ executes, and the \textbf{Static propensities} invariant holds in $S_1$.

SSA object $S_2$ will receive a \textit{Populations} message at the time $x$ executes, which will cause $S_2$ to cancel its pending reaction and redo reaction scheduling, as performed by lines 5--8 of the \textsc{\textit{Populations}} event method.
Therefore, the \textbf{Static propensities} invariant holds in $S_2$ as well, and the parallel SSA algorithm handles reactions that transfer species between compartments.

Since each reaction execution is independent, this analysis extends to multiple exchange reactions between a pair of adjacent compartments, and to many compartments.

Lastly, we note that an exact sequential SSA simulation of a system containing multiple compartments must use an approach similar to ours because it will simulate each compartment independently and cannot be exact unless it cancels reactions which are interrupted by exchange reactions.
However, unlike this parallel algorithm, it will not need to handle cascading cancellations.

\subsection{Practical considerations for the PDES simulator}
\label{section:Practical considerations for the PDES simulator}
The parallel SSA algorithm application employs several features that must be supported by the OO PDES simulator on which it runs.
First, because \textit{Populations} is a \textit{zero-delay} message \cite{jefferson2017virtual,jefferson2020virtual_time_III} these must be allowed by the simulator.
Second, the \textsc{Cancel scheduled reaction events} function uses a ``retract \textit{Event message}'' operation to send a \textit{retraction} message that cancels a previously sent \textit{Event message}.
Retraction messages.~\cite{lomow1991mechanisms,jefferson2020virtual_time_III} are supported by the ROSS optimistic PDES simulator \cite{Carothers2000,schordan2016automatic}

\section{Optimizations and evaluation}
\subsection{Optimizing parallel SSA}
\label{section:Optimizing parallel SSA}
Multiple optimizations should be incorporated into the parallel SSA algorithm to improve its performance.
These are all exact.
\begin{itemize}
  \item Multiple SSP objects should be employed so that a) they can share computational load, and b) shared species can be mapped to SSP objects that are located on a parallel computer near the SSA objects that use them.
  \item The optimization introduced by NRM that recomputes only propensities which depend on species populations that have changed should be incorporated.~\cite{gibson2000efficient}
  In parallel SSA, when a reaction executes these species are given by the union of the species with non-zero stoichiometry in the reaction with the shared species whose populations have been updated.
  When a reaction is cancelled, they are given by the updated species provided by the \textit{Populations} message that triggered the cancellation.
  This optimization should be implemented using the indexed priority queue $P$ defined in.~\cite{gibson2000efficient}
  This requires that SSPs track the shared species population updates which have not been received by each SSA.
  \item Unnecessary data in \textit{Populations} messages can be avoided by having SSA objects record updates to shared species locally and having SSPs not send an update to a shared species back to the SSA that reported the update.
  If all data in a \textit{Populations} message is unnecessary, the message need not be sent.
\end{itemize}

These optimizations are all straightforward to implement.

In addition to these exact optimizations, approximate optimizations should also be considered. 

\subsection{Evaluation}
While quantitative performance results are not available because parallel SSA has not been implemented yet, the parallel SSA algorithm achieves these conceptual objectives.
It offers a method to accelerate SSA without sacrificing accuracy by partitioning reaction networks and simulating sub-networks in parallel on a supercomputer.
In addition, it aims to reduce simulation run-times by maximizing the number of sub-networks in a simulation while minimizing the cost of synchronization.

The parallel SSA algorithm also faces several challenges.
Partitioning will be performed on a static reaction network, whereas the characteristics of the network and its sub-networks will likely vary over the duration of a simulation.
The performance of a parallel SSA simulation will depend on the rate of updates to shared species, and fraction of those updates that cause reactions to be cancelled. A partitioning algorithm that accurately estimates these rates needs to be developed.

\section{Next steps}
Much additional work must be completed before parallel SSA can become a standard tool for accelerating the simulation of large biochemical models.
We plan these next steps.

\begin{itemize}
  \item Implement the algorithm: Select a PDES simulator to use as a foundation, and implement the SSA algorithm and the optimizations from section~\ref{section:Optimizing parallel SSA} as an DES application that runs on the simulator.
  \item Implement partitioning: Develop a reaction-network partitioning algorithm.
  In addition, a related algorithm is needed to map SSA and SSP objects to processors and cores in a supercomputer when a parallel SSA simulation in initialized.
  \item Evaluate the implementation's performance: We will develop a benchmark reaction network, obtain a fast sequential SSA implementation such as \cite{somogyi2015libroadrunner}, and evaluate parallel SSA's relative speedup.
  Initial construction has begun on a configurable generator of synthetic reaction networks.
  \item Integrate parallel SSA into a user-friendly modeling environment: a comprehensive environment for modeling whole-cells and other large networks needs a modeling language, a simulation experiment language, and a format for simulation results.~\cite{goldberg2018emerging}
  \item Combine parallel SSA with other integration algorithms in a multi-algorithmic simulator: because different pathways in cells are characterized at different levels, whole-cell models must be simulated with multiple integration algorithms, including dynamic Flux Balance Analysis and ODEs along with SSA.~\cite{goldberg2018emerging}
  To achieve this, we will merge parallel SSA with an existing multi-algorithmic simulator.
\end{itemize}


\section{Conclusions}
We make important progress toward using parallelism to accelerate the Stochastic Simulation Algorithm (SSA) by presenting the first \textit{exact} parallel algorithm for SSA.
The algorithm parallelizes SSA with no loss of accuracy by partitioning a reaction network into multiple sub-networks that are simulated by independent but coordinated SSA instances.
It exactly synchronizing accesses to species populations shared by the instances, and cancels pending reactions that are interrupted by population updates that invalidate prior propensity calculations.
To recover from reaction cancellations the algorithm employs a modified Next Reaction Method \cite{gibson2000efficient} approach.
All concurrent synchronization, including event timing, reaction cancellation and rollback, is achieved by leveraging the existing synchronization in optimistic parallel DES simulators.
This will make the parallel SSA algorithm easier to implement and deploy.
In addition, the algorithm exactly simulates systems that contain multiple compartments and transfer species between them.
We present a plan for implementing, optimizing and evaluating it.

A high-performance, production parallel SSA algorithm would help enable simulations of comprehensive models of the entire biochemistry of cells, which would advance the treatment of disease and the engineering of useful microbes.

\begin{acknowledgments}
This worked was supported by National Science Foundation award 1649014 and National
Institutes of Health award R35GM119771 to J.R.K.
We appreciate insightful comments from Robert Clayton Blake at Lawrence Livermore National Laboratory.
\end{acknowledgments}


  \bibliography{parallel-SSA}
\end{document}